\documentclass[english,numberedappendix]{emulateapj}
\usepackage[T1]{fontenc}
\usepackage{graphicx}

\makeatletter
\slugcomment{}
\shorttitle{Models of Bursts with Short Recurrence Times}
\shortauthors{Keek \& Heger}

\makeatother

\usepackage{babel}
\begin{document}

\title{Thermonuclear Bursts with Short Recurrence Times from Neutron Stars
Explained by Opacity-Driven Convection}

\author{L.~Keek\altaffilmark{1,2} \& A.~Heger\altaffilmark{3,4,5}}

\altaffiltext{1}{X-ray Astrophysics Laboratory, Astrophysics Science Division, NASA/GSFC, Greenbelt, MD 20771} 
\altaffiltext{2}{CRESST and the Department of Astronomy, University of Maryland, College Park, MD 20742}
\altaffiltext{3}{Monash Center for Astrophysics, School of Physics and Astronomy, Monash University, Victoria, 3800, Australia}
\altaffiltext{4}{Shanghai Jiao-Tong University, Center for Nuclear Astrophysics, Department of Physics and Astronomy, Shanghai 200240, P.R. China}
\altaffiltext{5}{University of Minnesota, School of Physics and Astronomy, Minneapolis, MN 55455}

\email{laurens.keek@nasa.gov}
\begin{abstract}
Thermonuclear flashes of hydrogen and helium accreted onto neutron
stars produce the frequently observed Type I X-ray bursts. It is the
current paradigm that almost all material burns in a burst, after
which it takes hours to accumulate fresh fuel for the next burst.
In rare cases, however, bursts are observed with recurrence times
as short as minutes. We present the first one-dimensional multi-zone
simulations that reproduce this phenomenon. Bursts that ignite in
a relatively hot neutron star envelope leave a substantial fraction
of the fuel unburned at shallow depths. In the wake of the burst,
convective mixing events driven by opacity bring this fuel down to
the ignition depth on the observed timescale of minutes. There, unburned
hydrogen mixes with the metal-rich ashes, igniting to produce a subsequent
burst. We find burst pairs and triplets, similar to the observed instances.
Our simulations reproduce the observed fraction of bursts with short
waiting times of $\sim30\,\%$, and demonstrate that short recurrence
time bursts are typically less bright and of shorter duration. 
\end{abstract}


\section{Introduction}

\label{sec:Introduction}

Type I X-ray bursts are frequently observed from neutron stars that
accrete through Roche-lobe overflow from a lower-mass companion star
\citep[for reviews, see][]{Lewin1993,Strohmayer2006,Galloway2008catalog}.
The material accumulated on the neutron star surface is rich in helium
and also hydrogen in most cases. Runaway thermonuclear burning in
this layer produces an X-ray flash that typically lasts $10-100\,\mathrm{s}$.
It is the current paradigm that most of the accreted fuel is burned
during a burst, with only a sliver of fuel remaining on top \citep[e.g.,][]{Woosley2004,Fisker2008,Jose2010}.
The atmosphere must, therefore, be almost completely replaced by accretion
before a new burst can ignite (see \citealt{Woosley2004} for the
effect of compositional inertia). Depending on the mass accretion
rate, $\dot{M}$, X-ray bursts are observed to recur on timescales
of hours, days, or even longer. An exceptional source is IGR~J17480$-$2446,
which has exhibited regularly repeating bursts with recurrence times
as short as $t_{\mathrm{recur}}\simeq3.3\,\mathrm{min}$ \citep{Motta2011,Linares2011}.
There are, however, instances where the regularity is broken, and
the distribution of $t_{\mathrm{recur}}$ is bimodal. At a constant
$\dot{M}$, recurrence times of both a few hours and a few minutes
appear. 

Using the nomenclature of previous publications, we distinguish short
waiting time (SWT) and long waiting time (LWT) bursts, depending on
whether the time since the previous burst was shorter or longer than
$45$~minutes, respectively \citep{Boirin2007,Keek2010}. LWT bursts
have been observed to be followed by one \citep{1608:murakami80pasj,0748:gottwald87apj},
two \citep{Boirin2007}, and even three SWT bursts \citep{Galloway2008catalog,Keek2010}.
Such events are referred to as double, triple, and quadruple bursts.
Even when a source is in a state where it displays SWT bursts, many
LWT bursts are not followed by an SWT burst (``single bursts'').
The occurrence of SWT bursts appears to be random, with a $\sim30\%$
probability for each burst (both LWT and SWT) to be followed by an
SWT burst \citep{Boirin2007,Keek2010}. This suggests that a stochastic
process is important for the ignition of SWT bursts.

Short recurrence times have been observed in rare cases from $15$
sources \citep{Keek2010}. All sources are hydrogen-accretors with
relatively high spin frequencies of $\nu\ge549\,\mathrm{Hz}$. For
individual sources it was found that SWT bursts are restricted to
a range of $\dot{M}$, although the range does not always overlap
for all sources, which may be due to the uncertainties in determining
$\dot{M}$ for each source. SWT bursts are typically less bright and
less energetic than LWT bursts. Also, their shorter duration and the
shape of their light curve indicate that SWT bursts are powered by
fuel with a reduced hydrogen content compared to LWT bursts \citep{Boirin2007}.

With the mentioned exception of IGR~J17480$-$2446, recurrence times
of minutes are too short to accrete the fuel of an SWT burst. These
bursts must be powered by fuel left-over from the previous burst.
It has long been unexplained, however, how and where this fuel is
preserved. The blackbody radius measured for the SWT and LWT bursts
are consistent with being the same \citep{Boirin2007}, which disfavors
explanations of this phenomenon that call on the burning of separate
patches on the neutron star surface. Alternatively, fuel may be separated
in layers at different depths in the atmosphere \citep{1636:fujimoto87apj}.
A second part of this puzzle is the reignition of this fuel on a short
timescale. The short recurrence times vary between $3\,\mathrm{min}$
and $45\,\mathrm{min}$ \citep{Keek2010}: there is a considerable
spread, even for SWT bursts from a single source, so it is unlikely
to be related to a nuclear waiting point in the \textsl{$\alpha$p}-
and \textsl{rp}-processes, which would have a sharply defined time
scale \citep{Boirin2007}.

In this paper we present one-dimensional models of the neutron star
envelope, where SWT bursts are produced in a self-consistent manner
with properties that closely reproduce many of the observed features.
After introducing our numerical implementation (Section~\ref{sec:Neutron-star-envelope}),
we show models where a substantial fraction of the fuel survives the
LWT burst, and is transported by convection to the ignition depth,
producing an SWT burst on the observed timescale (Section~\ref{sec:Results}).
We compare the models to the observed SWT events, and discuss the
accuracy of our convection implementation (Section~\ref{sec:Discussion}),
before presenting our conclusions (Section~\ref{sec:Conclusions}).

\section{Neutron Star Envelope Model}

\label{sec:Neutron-star-envelope}

We employ the successful burst model implemented in the hydrodynamics
stellar evolution code \textsc{Kepler} \citep{Weaver1978,Woosley2004,Keek2011,Keek2014}.
Here we describe the main properties of the model, and we refer to
the cited papers for further details. An adaptive one-dimensional
Lagrangian grid describes the outer layers of the neutron star in
the radial direction. The base is formed by a $10^{25}\,\mathrm{g}$
iron layer, on top of which material is accreted of solar composition
(mass fractions of $0.71$ $^{1}$H, $0.27$ $^{4}$He, and $0.02$
$^{14}$N), since all observed SWT bursts are from hydrogen-rich sources
\citep{Keek2010}. The outer zone has a mass of $10^{16}\,\mathrm{g}$.
Accretion is implemented by using a finely resolved grid in the outer
layers that is fixed relative to the surface. Advection of composition
as well as compressional heating due to accretion is taken into account.
Below a specified mass, which is chosen such that most nuclear reactions
occur at greater depth, a fully Lagrangian grid is used \citep{Keek2011}.
In the present models, the specified mass corresponds to a column
depth of $8\times10^{6}\,\mathrm{g}\,\mathrm{cm}^{-2}$.

We report the accretion rate of our simulations as a fraction of the
Eddington-limited rate for solar composition: $\dot{M}_{\mathrm{Edd}}=1.75\times10^{-8}\,M_{\odot}\mathrm{yr^{-1}}$.
Nuclear reactions are followed using a large adaptive network (\citealt{Rauscher2000,Rauscher2002})
that includes the (``hot'') $\beta$-limited CNO cycle, the $3\alpha$,
\textsl{$\alpha$p}-, and \textsl{rp}-processes (\citealt{Wallace1981}).
Neutrino losses from weak decays are taken into account. Chemical
mixing due to convection uses mixing-length theory with a mixing-length
parameter of $\alpha=1$ and is implemented as a diffusive process
(e.g., \citealt{Clayton1968book}). Semiconvection is modeled as a
diffusive process with a diffusion coefficient that is $10\,\%$ of
that of thermal diffusion, similar in efficiency to the formulation
by \citet{Langer1983} with an $\alpha$ value of $0.04$. Thermohaline
mixing is modeled following \citet{Heger2005a}, but mostly plays
a minor role in the ashes layers.

Radiative opacity is implemented using analytic fits by \citet{Iben1975}
to numerical opacity calculations by \citet{Cox1970a,Cox1970b}, including
electron scattering, Compton scattering, bound-free, and free-free
transitions \citep[see also][]{Weaver1978}. Furthermore, the opacity
due to electron conduction follows \citet{Itoh2008}.

Pycno-nuclear as well as electron-capture reactions in the crust produce
a heat flux into the neutron star envelope \citep{Haensel1990,Haensel2003,Gupta2007}.
An unknown shallow heat source may add to this \citep{Brown2009,Deibel2015,Turlione2015},
whereas neutrino cooling by Urca cycling could reduce it \citep{Schatz2013,Deibel2016}.
We calculate models for a range of values of the heating per accreted
nucleon, $Q_{\mathrm{b}}$ , in units of $\mathrm{MeV\,u^{-1}}$,
with $u$ the atomic mass unit, and set the luminosity at the inner
boundary to $L_{\mathrm{b}}=\dot{M}Q_{\mathrm{b}}$.

The local gravity in our model is the Newtonian value for a $1.4\,M_{\odot}$
neutron star with a radius of $R=10\,\mathrm{km}$. The results presented
in this paper are in the reference frame of the model domain. They
can be corrected for General Relativistic effects \citep{Keek2011}:
keeping the gravitational mass of $1.4\,M_{\odot}$, the same surface
gravity is obtained including GR by increasing the stellar radius
to $R=11.2\,\mathrm{km}$, which produces a gravitational redshift
of $z=0.26$ (e.g., \citealt{Woosley2004}). For example, an observer
at infinity will measure a $t_{\mathrm{recur}}$ that is $26\,\%$
longer.

\section{Results}

\label{sec:Results}

\subsection{Simulations for a Range of Base Heating}

\begin{figure}
\includegraphics{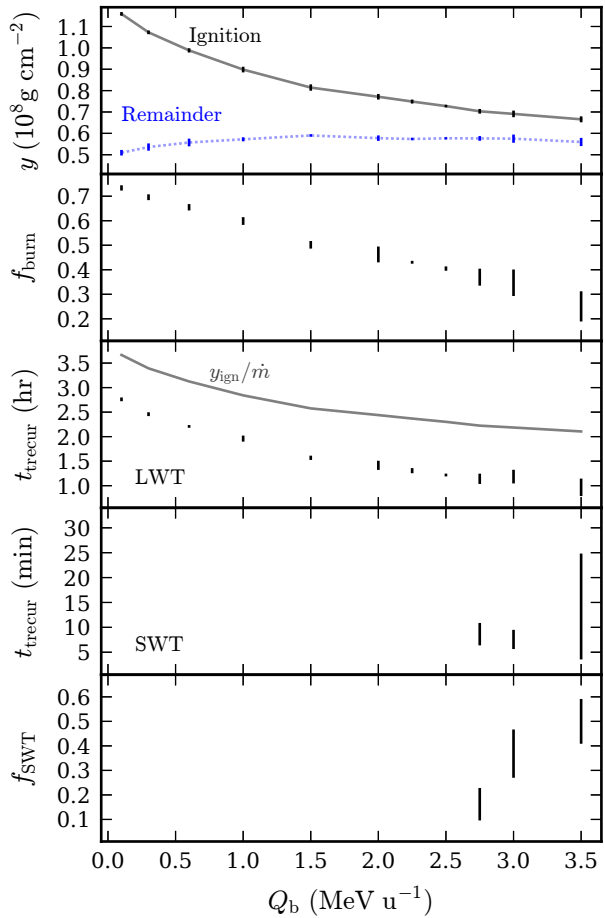}

\caption{\label{fig:qb_overshoot}Properties of the bursts for a series of
simulations with $\dot{M}=0.1\,\dot{M}_{\mathrm{Edd}}$ as a function
of the choice of base heating, $Q_{\mathrm{b}}$. Shown are the column
depths, $y$, where the LWT bursts ignite and where burning stops,
leaving a remainder of unburned fuel; the fraction of the fuel burned
in the LWT bursts, $f_{\mathrm{burn}}$; the recurrence times, $t_{\mathrm{recur}}$,
of the LWT and the SWT bursts, where for the former a solid line indicates
the predicted values if all fuel had burned; the fraction of bursts
that have a short waiting time, $f_{\mathrm{SWT}}$, which is zero
for all but the three largest values of $Q_{\mathrm{b}}$. The error
bars indicate the root mean squared variation of the quantities for
a series of bursts in one simulation.}
\end{figure}
 
\begin{figure*}
\includegraphics{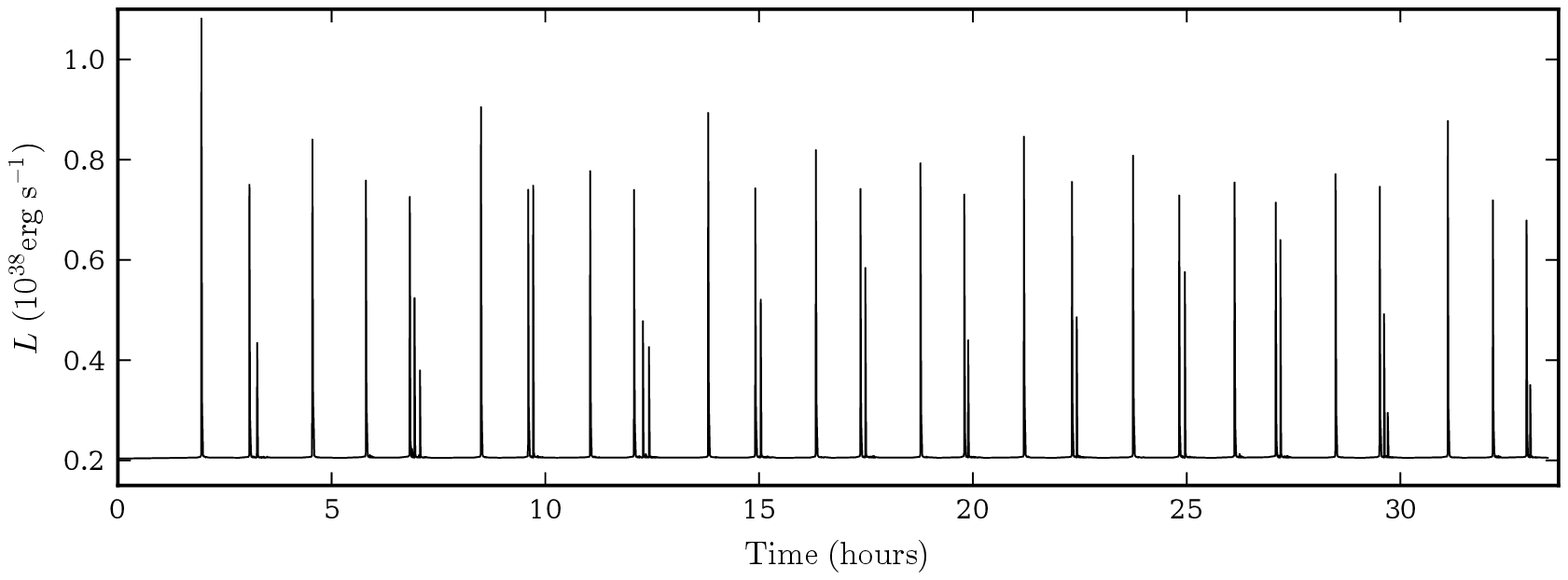}

\caption{\label{fig:lcv_xh15}Both LWT and SWT bursts are visible in the light
curve of a simulation with  $\dot{M}=0.1\dot{M}_{\mathrm{Edd}}$
and $Q_{\mathrm{b}}=3.0\,\mathrm{MeV\,u^{-1}}$. The accretion luminosity
of $0.2\times10^{38}\,\mathrm{erg\,s^{-1}}$ is included. The qualitative
bursting behavior is strikingly similar to that seen during long observations
of EXO~0748-676 \citep{Boirin2007,Keek2010}.}
\end{figure*}

We create $11$ simulations with $\dot{M}=0.1\,\dot{M}_{\mathrm{Edd}}$
for values of $Q_{\mathrm{b}}$ ranging from $0.1\,\mathrm{MeV\,u^{-1}}$
to $3.5\,\mathrm{MeV\,u^{-1}}$. The simulations produce a series
of on average $22$ LWT bursts, and at the highest three values of
$Q_{\mathrm{b}}$ both SWT and LWT bursts appear. The recurrence times
of the two groups are sufficiently different to clearly distinguish
the SWT from the LWT bursts (Figure~\ref{fig:qb_overshoot}). The
shortest waiting time before an SWT burst is $4.6\,\mathrm{min}$.
The fraction of bursts that are SWT, $f_{\mathrm{SWT}}$, increases
with $Q_{\mathrm{b}}$. 

We determine the properties of the bursts in each simulation, excluding
the first and the last burst. The recurrence time is measured as the
time between the peaks in the luminosity of subsequent bursts; the
fluence is measured by integrating the luminosity starting when it
first exceeds $25\%$ of the peak value up to the time it drops below
$10^{36}\,\mathrm{erg\,s^{-1}}$. The size of the fuel column is determined
by the location where the hydrogen mass fraction is reduced to $0.1$
(compared to the accreted value of $0.71$), and we give the value
at the start and the end of the burst (``ignition'' and ``remainder''
in Figure~\ref{fig:qb_overshoot}).

For all simulations, only part of the fuel column is burned by LWT
bursts (Figure~\ref{fig:qb_overshoot}). For the coldest model, with
$Q_{\mathrm{b}}=0.1\,\mathrm{MeV\,u^{-1}}$, a fraction of $f_{\mathrm{burn}}=0.73$
of the fuel is burned, whereas for the largest $Q_{\mathrm{b}}$ a
mere $25\,\%$ is burned. The ignition column depth of the LWT bursts
is smaller for larger $Q_{\mathrm{b}}$ and ranges from $1.16\times10^{8}\,\mathrm{g\,cm^{-2}}$
to $0.67\times10^{8}\,\mathrm{g\,cm^{-2}}$. The unburned remaining
column exhibits a much smaller variation: for larger $Q_{\mathrm{b}}$
it increases from $0.51\times10^{8}\,\mathrm{g\,cm^{-2}}$ to $0.56\times10^{8}\,\mathrm{g\,cm^{-2}}$:
whereas the relative fraction of unburned fuel increases with $Q_{\mathrm{b}}$,
the absolute amount of remaining fuel is similar in all cases. An
important consequence is that the remaining fuel is closer to the
ignition depth for larger $Q_{\mathrm{b}}$. 

If the entire fuel column needs to be replaced by accretion, the recurrence
time is simply $t_{\mathrm{recur}}=y_{\mathrm{ign}}/\dot{m}$, with
the specific mass accretion rate $\dot{m}=\dot{M}/4\pi R^{2}$ (in
the Newtonian frame). We find, however, substantially shorter recurrence
times even for the LWT bursts, as only the burned part of the column
needs to be replaced by accretion (Figure~\ref{fig:qb_overshoot}).

\subsection{Comparison of LWT and SWT Bursts}

\label{subsec:Comparison-of-LWT}

For $Q_{\mathrm{b}}=3.0\,\mathrm{MeV\,u^{-1}}$, $f_{\mathrm{SWT}}=0.37\pm0.10$,
which is consistent with the value of $\sim0.3$ derived from observations
\citep{Boirin2007,Keek2010}. The light curve of this simulation (Figure~\ref{fig:lcv_xh15})
looks qualitatively similar to the long exposures of the \emph{XMM-Newton}
and \emph{Chandra} observatories on EXO~0748-676 \citep{Boirin2007,Keek2010},
with the appearance of burst doublet and triplet events seemingly
at random. We employ this simulation to study the properties of the
$24$ LWT and $14$ SWT bursts in more detail. 

\begin{figure}
\includegraphics{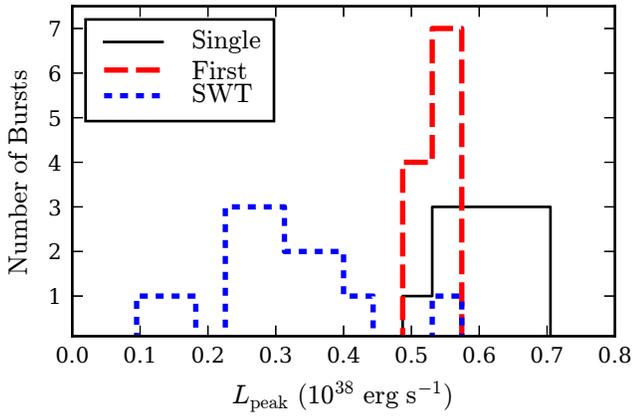}

\caption{\label{fig:hist_peaklum} Histogram of the peak luminosity, $L_{\mathrm{peak}}$,
for single bursts, the first bursts of multiple events, and SWT bursts
from Figure~\ref{fig:lcv_xh15}. The SWT bursts are typically less
bright than the LWT bursts. }
\end{figure}
 
\begin{figure}
\includegraphics{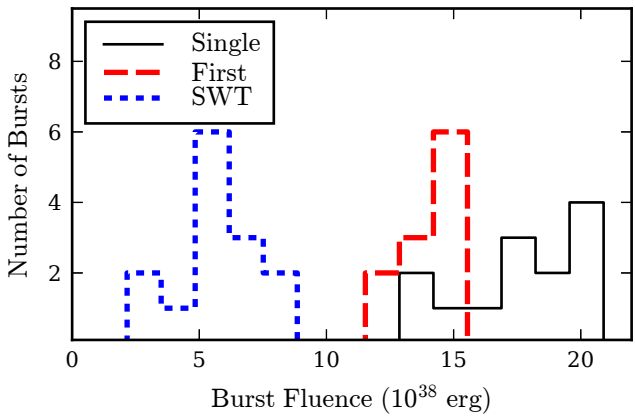}

\caption{\label{fig:hist_burst_fluence} Histogram of the burst fluence for
the same burst categories as Figure~\ref{fig:hist_peaklum}. The
SWT bursts are less energetic than the LWT bursts. }
\end{figure}
 
\begin{figure}
\includegraphics{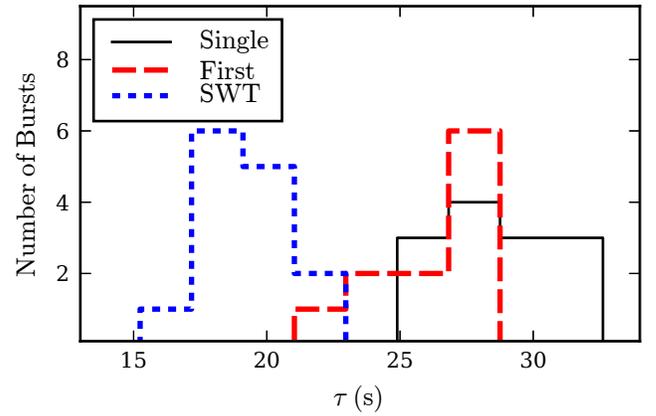}

\caption{\label{fig:hist_tau} Histogram of the decay time scale, $\tau$,
(the ratio of the burst fluence to the peak luminosity) for the same
burst categories as Figure~\ref{fig:hist_peaklum}. The SWT bursts
last shorter than the LWT bursts.}
\end{figure}
The bursts are divided among three categories: ``single'' bursts
(LWT not followed by an SWT burst), ``first'' bursts (LWT followed
by SWT), and SWT bursts. For each burst we determine the fluence,
peak luminosity, $L_{\mathrm{peak}}$, and the ratio of the two, which
gives the decay timescale, $\tau$. The SWT bursts typically have
a lower $L_{\mathrm{peak}}$ (Figure~\ref{fig:hist_peaklum}) and
fluence (Figure~\ref{fig:hist_burst_fluence}) as well as a shorter
$\tau$ (Figure~\ref{fig:hist_tau}). The distributions of these
properties largely overlap for the single and first bursts. Interestingly,
however, the highest values originate with single bursts. It indicates
that when more fuel is consumed in an LWT burst, the subsequent appearance
of an SWT burst is less likely.

The $\alpha$-parameter is the ratio of the persistent fluence since
the previous burst (including the accretion fluence) to the burst
fluence. For the LWT bursts, we find values in the range of $47$
to $66$, with no notable difference in the distributions for single
and first bursts. The SWT bursts have on average much lower $\alpha$
values ranging from $13$ to $42$.

\begin{figure}
\includegraphics{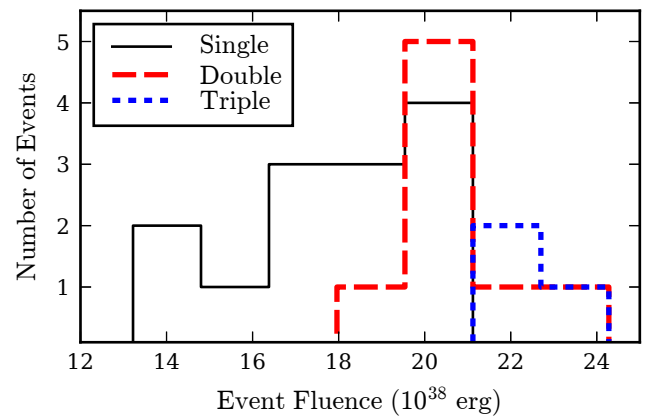}

\caption{\label{fig:hist_event_fluence} Histogram of the fluence of burst
events, i.e., where we sum the fluence of two bursts in a double or
three in a triple event. The distributions for double and triple are
consistent with the highest values for single, which suggests that
SWT bursts are powered by left-over fuel.}
\end{figure}
 When summing the fluence of each LWT burst and any following SWT
bursts, we see that the combined fluence in double and triple events
is as large as the most energetic single bursts (Figure~\ref{fig:hist_event_fluence}),
which is also an indication of the burning of left-over fuel in SWT
bursts.

As mentioned, the properties of the LWT and SWT bursts in our simulation
are qualitatively similar to those observed from EXO~0748-676 \citep[see Figure 7 of][]{Boirin2007}
as well as a large compilation of observations of bursting sources
\citep[compare with Figures 9-12 of][]{Keek2010}. For the latter,
the combination of multiple sources and a wide range of mass accretion
rates produces histograms with broader distributions than our simulation
at a constant mass accretion rate. Nonetheless, we see the same behavior,
with SWT bursts being on average less bright and of shorter duration
than LWT bursts, and the summed fluence in multiple-burst events being
comparable to the fluence of the more energetic single bursts.

\subsection{Detailed Look at a Burst Triplet}

To investigate the processes that produce SWT bursts, we study one
triple burst in detail, which we take from the simulation with $Q_{\mathrm{b}}=3.0\,\mathrm{MeV\,u^{-1}}$
(see at $6.7\,\mathrm{hr}$ in Figure~\ref{fig:lcv_xh15}).
\begin{figure}
\includegraphics{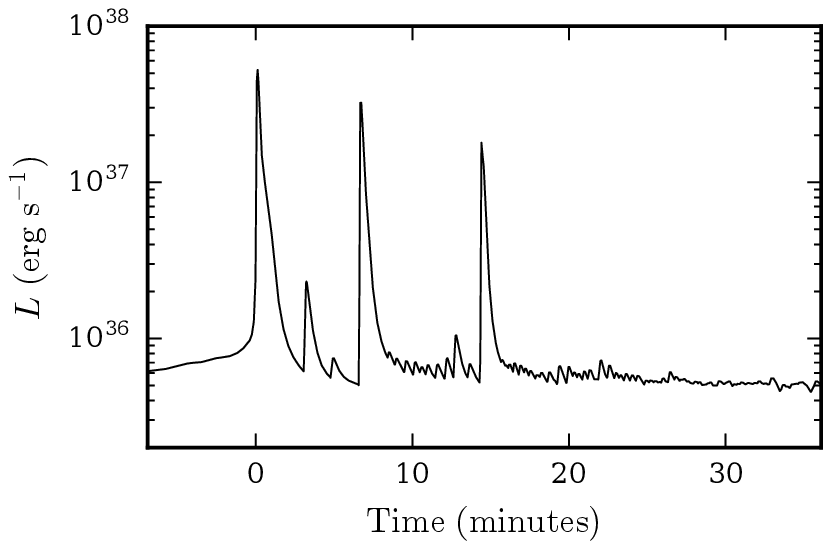}

\caption{\label{fig:lcv_xh15_zoom}The light curve of the triple burst event
at $6.7\,\mathrm{hr}$ in Figure~\ref{fig:lcv_xh15} (the accretion
flux is not included here). Aside from the three bursts, smaller bumps
are visible due to convective mixing events. }
\end{figure}
 Its light curve exhibits $3$ bursts as well as a series of smaller
bumps (Figure~\ref{fig:lcv_xh15_zoom}). The bumps originate from
mixing events near the burst ignition depth, and involve a small amount
of localized burning. In contrast, nuclear burning is stronger during
the bursts, spreading from the ignition location to smaller depth,
and powering a flare that is over an order of magnitude brighter than
a bump. With a peak luminosity of $\lesssim10\,\%$ of the accretion
luminosity, it will be challenging to detect these bumps in X-ray
observations (Figure~\ref{fig:lcv_xh15}).

\begin{figure}
\includegraphics{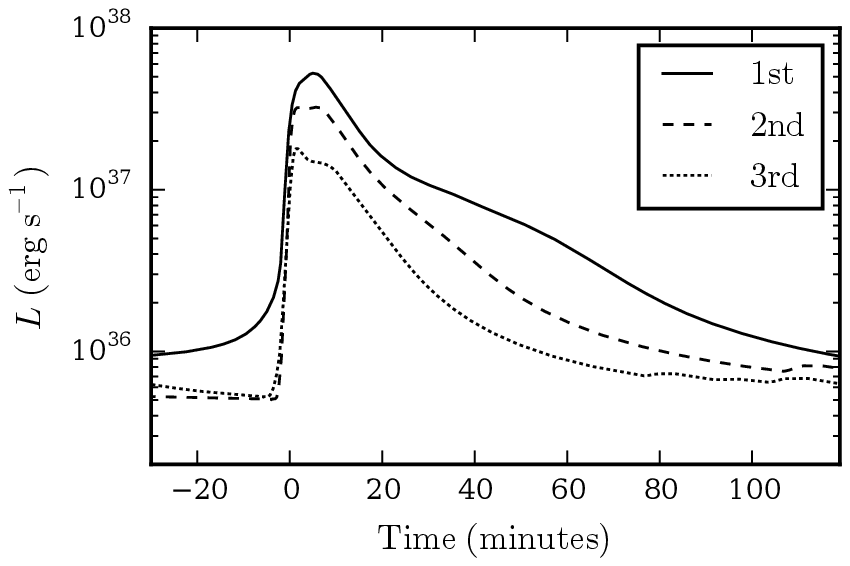}

\caption{\label{fig:lcv_xh15_zoom1}Comparison of burst profiles of the triple
burst in Figure~\ref{fig:lcv_xh15_zoom}. The light curves have been
shifted in time such that their rises coincide at $0$.}
\end{figure}
 Comparing the light curves of the three bursts, we see that the first
burst is the brightest with a long tail that shows a substantial contribution
from the \textsl{rp}-process, as evidenced by the ``hump'' around
$\sim50\,\mathrm{s}$ (Figure~\ref{fig:lcv_xh15_zoom1}). The subsequent
SWT bursts have a lower $L_{\mathrm{peak}}$, a smaller fluence, and
a shorter duration, which we found to be typical for SWT bursts in
our simulations (Section~\ref{subsec:Comparison-of-LWT}).

\subsubsection{Incomplete burning}

\begin{figure}
\includegraphics{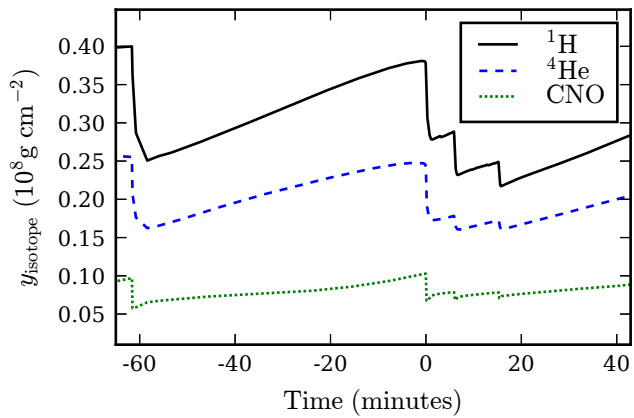}

\caption{\label{fig:lcv_xh15_isotopes}Total mass column of hydrogen, helium,
and the combined CNO isotopes for the triple burst in Figure~\ref{fig:lcv_xh15_zoom}.
After the previous burst on the left-hand side, accretion slowly increases
the fuel column over time, whereas the three bursts cause sudden decreases.
After each burst in the triplet, accretion does not fully compensate
for the burned fuel burned before the next burst ignites.}
\end{figure}
 The waiting time before the LWT burst was $1.03\,\mathrm{hr}$, during
which a column of $\Delta y=0.32\times10^{8}\,\mathrm{g\,cm^{-2}}$
was accreted. Burst ignition occurs deeper at $y_{\mathrm{ign}}=0.65\times10^{8}\,\mathrm{g\,cm^{-2}}$:
material that was accreted prior to the preceding burst is pushed
deeper by the fresh fuel on top, and this material now ignites a burst.
We track the total column of hydrogen, helium, and CNO in the model
as a function of time (Figure~\ref{fig:lcv_xh15_isotopes}). After
the first burst, a substantial fraction of hydrogen and helium remains.
Accretion slowly increases their columns over time, but this increase
is minor in the short waiting time before the SWT bursts. Reignition
involves, therefore, left-over fuel instead of fresh fuel. The amount
of fuel burned is smaller for each successive burst. The amount of
helium burned in the SWT bursts is relatively small. Furthermore,
the small reduction in the CNO column during the SWT bursts, is quickly
compensated by $3\alpha$ burning after the bursts.

\begin{figure}
\includegraphics{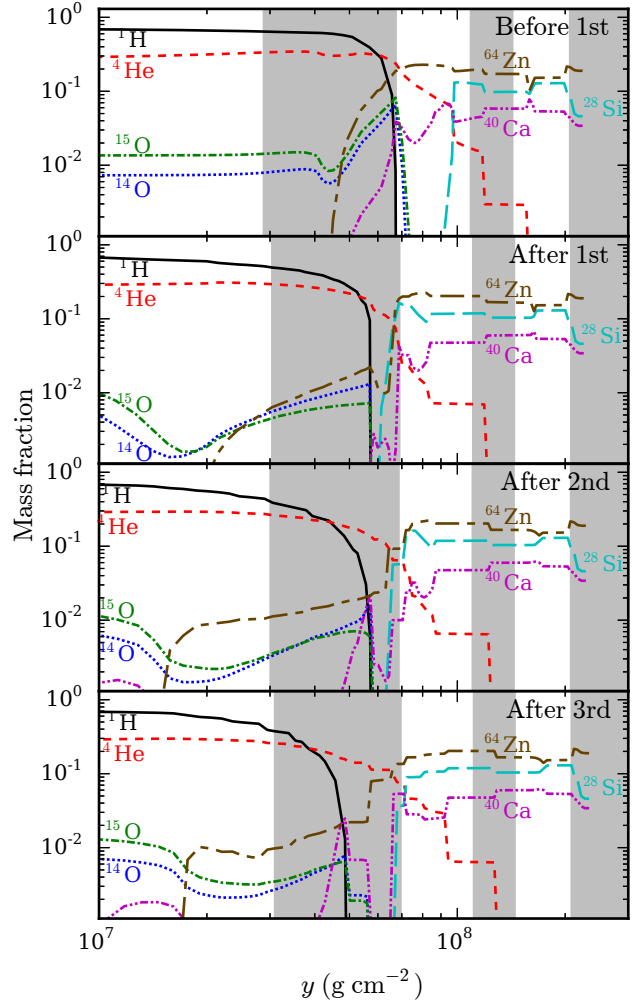}

\caption{\label{fig:Composition-of-the}Composition of the envelope $1\,\mathrm{min}$
before the burst triplet and $2\,\mathrm{min}$ after each of the
three bursts. Shown are the mass fractions as a function of column
depth, $y$, for a small selection of the most abundant isotopes.
Alternate white and gray shaded areas correspond to the layers accreted
between subsequent LWT bursts, and the small offsets visible between
the panels are due to small amounts of accretion between the shown
instances. At large depths (on the right-hand side) is the outer part
of the iron substrate of the model.}
\end{figure}
 We take a closer look at the composition as a function of depth at
several times: before the LWT burst and after each of the three bursts
(Figure~\ref{fig:Composition-of-the}). Ignition of the LWT burst
takes place close to the bottom of the hydrogen column, which is also
close to the bottom of the layer that was accreted before the \emph{previous}
LWT burst. Furthermore, the ignition location contains ashes of the
previous burst. The most abundant heavy isotope is $^{64}$Zn, which
is produced by the \textsl{rp}-process, and some $\alpha$ elements
such as $^{40}$Ca are present with smaller mass fractions. These
ashes produce so-called compositional inertia \citep{Woosley2004},
which reduces the recurrence time of bursts (e.g., in Figure~\ref{fig:lcv_xh15}
the waiting time before the first burst is longer than for the others).
Near $y_{\mathrm{ign}}$ the hydrogen mass fraction has been reduced
by $\beta$CNO burning before the LWT burst. After each subsequent
burst, the hydrogen column is further reduced. Moreover, the steepness
of the hydrogen profile is reduced, mostly by the mixing events in
between the bursts. The SWT bursts, therefore, are primarily powered
by hydrogen that is already available at the start of the LWT burst.

Before the LWT burst, the CNO abundance peaks at the bottom of the
hydrogen column (Figure~\ref{fig:Composition-of-the}). After this
burst, the CNO mass fractions are reduced at smaller column depths.
The lack of CNO inhibits hydrogen burning at smaller column depth,
facilitating its survival.

Each burst sends a heat wave into the deeper layers, inducing $\alpha$-captures
that produce isotopes such as $^{28}$Si \citep[see also][]{Woosley2004}.

\subsubsection{Reignition}

\begin{figure*}
\includegraphics[width=1\textwidth]{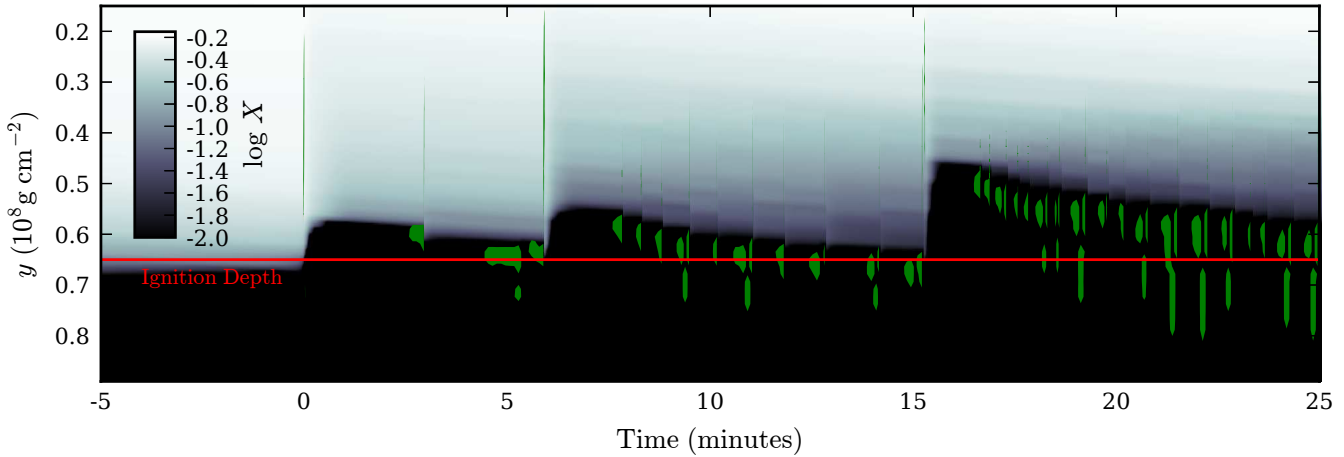}

\caption{\label{fig:h1_mixing} The hydrogen mass fraction, $X$, as a function
of time and column depth, $y$, during the burst triplet from Figure~\ref{fig:lcv_xh15_zoom}.
Accreted material is on top, and burst ashes are at the bottom. The
horizontal line indicates the ignition depth ($y_{\mathrm{ign}}$)
of the three bursts, and the green areas mark convective regions.
Mixing transports hydrogen in steps down to $y_{\mathrm{ign}}$, where
it ignites and powers the SWT bursts.}
\end{figure*}

Figure~\ref{fig:h1_mixing} presents a detailed view of the location
of hydrogen fuel during the burst triplet. After each burst a series
of brief convective mixing events occur. Convection takes place in
the ashes layer directly below the left-over hydrogen. Convective
overshooting extends the mixing region, dragging hydrogen down into
the ashes layer. This causes a small amount of burning and shuts off
convection. Subsequent mixing events bring hydrogen to larger column
depths. For all three bursts, ignition occurs near the same depth
$y_{\mathrm{ign}}=0.65\times10^{8}\,\mathrm{g\,cm^{-2}}$ (Figure~\ref{fig:h1_mixing}).
Once the convective events reach this depth, the mixed-in fuel ignites
as a new burst. 

\begin{figure*}
\includegraphics{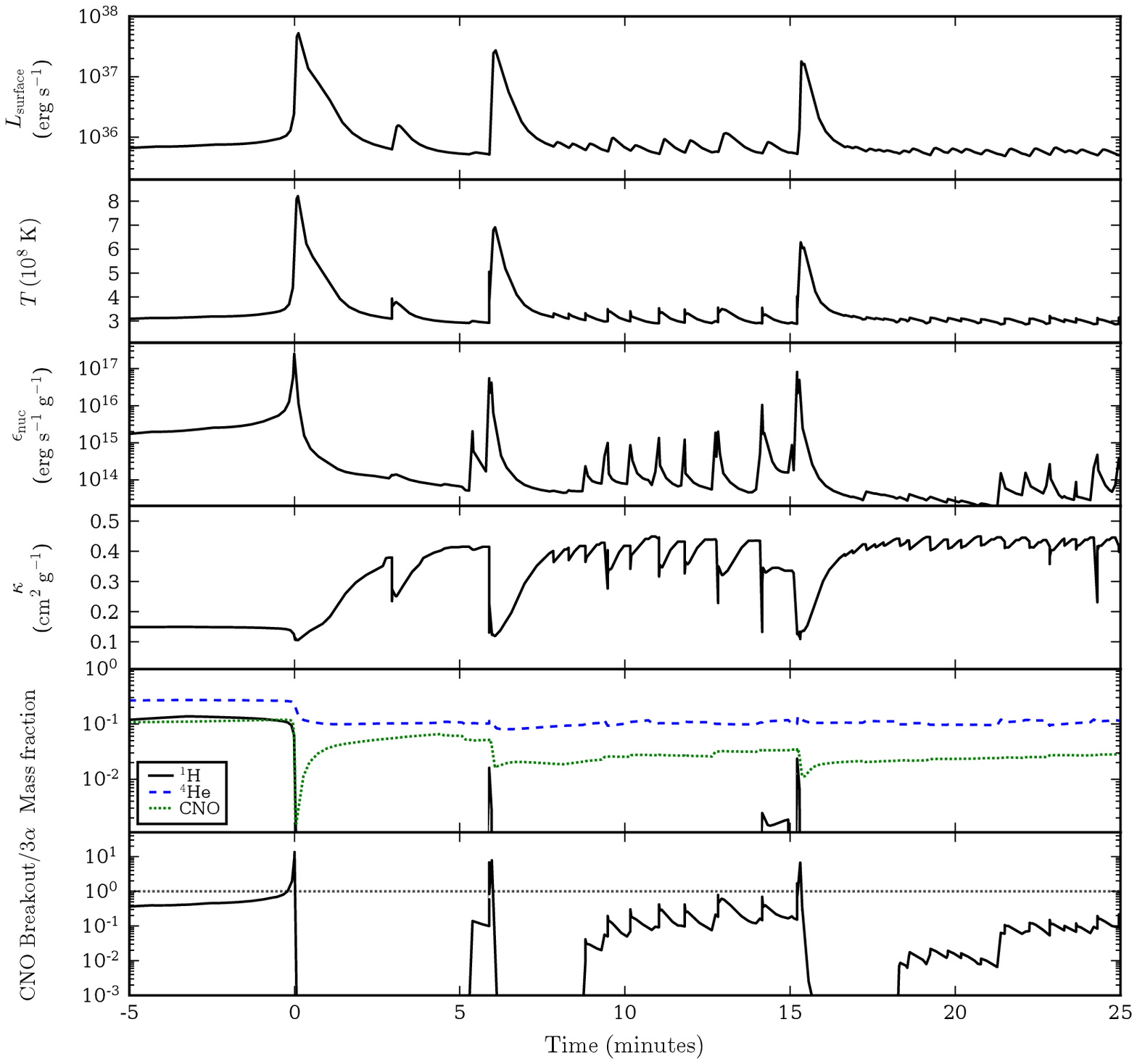}

\caption{\label{fig:flows_xh15}Properties of the triple burst event in Figure~\ref{fig:lcv_xh15_zoom}
as a function of time since the start of the first burst. The top
panel shows the surface luminosity $L_{\mathrm{surface}}$, whereas
the other panels show properties at the ignition depth of $y_{\mathrm{ign}}=0.65\times10^{8}\,\mathrm{g\,cm^{-2}}$:
temperature $T$, specific nuclear energy generation rate $\epsilon_{\mathrm{nuc}}$,
opacity $\kappa$, mass fractions of hydrogen helium and the combined
CNO elements, and the ratio of the nuclear flows through the $\beta$CNO
breakout reactions ($^{15}\mathrm{O}\left(\alpha,\gamma\right)\mathrm{^{19}Ne}$
and $^{18}\mathrm{Ne}\left(\alpha,p\right)\mathrm{^{21}Na}$) to the
$3\alpha$ reaction. In the bottom panel the dotted horizontal line
indicates unity, showing that only during the three bursts does breakout
flow exceed the $3\alpha$ flow.}
\end{figure*}
 In order to understand why the mixing events occur, consider the
evolution of several quantities at $y_{\mathrm{ign}}$ (Figure~\ref{fig:flows_xh15}).
During each burst, nuclear burning (indicated by the specific nuclear
energy $\epsilon_{\mathrm{nuc}}$) depletes hydrogen at this depth,
and strongly raises the temperature, $T$. This is followed by several
minutes of cooling, which lowers $T$ back down. The temperature and
composition changes have a strong effect on the opacity, $\kappa$.
Comparing $\kappa$ before and after the first burst shows the effect
of burning all hydrogen. The compositional changes are less radical
for the SWT bursts, but the temperature evolution induces changes
in $\kappa$ of similar magnitude. 
\begin{figure}
\includegraphics{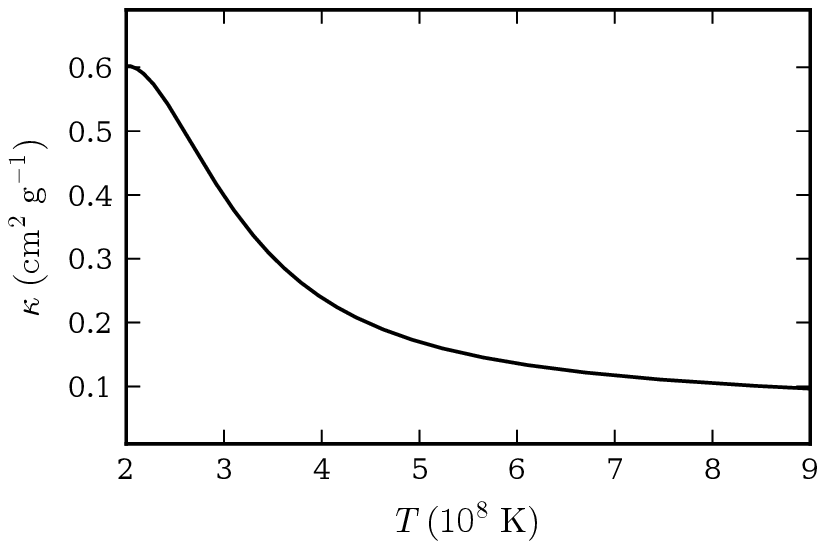}

\caption{\label{fig:kappa_t}Opacity, $\kappa$, as a function of temperature,
$T$. We take the model at $t=5\,\mathrm{minutes}$ and the zone at
$y_{\mathrm{ign}}$ (Figure~\ref{fig:flows_xh15}), which is devoid
of hydrogen and has a density of $\rho=5.3\times10^{5}\,\mathrm{g\,cm^{-3}}$,
and we calculate $\kappa$ in the range of $T$ that is relevant for
the triple burst (Figure~\ref{fig:flows_xh15}). The density dependence
is small: for this zone $\rho=(2.7-5.4)\times10^{5}\,\mathrm{g\,cm^{-3}}$,
which produces $\kappa=0.37-0.41\,\mathrm{cm^{2}\,g^{-1}}$.}
\end{figure}
 To illustrate the temperature dependence, we calculate $\kappa$
for a range of temperatures, using the properties of the model at
$y_{\mathrm{ign}}$ and $t=5$~minutes after the LWT burst (Figure~\ref{fig:kappa_t}).
Within the range of temperatures spanned in our triple burst, cooling
clearly produces an increase in the opacity. 
\begin{figure}
\includegraphics{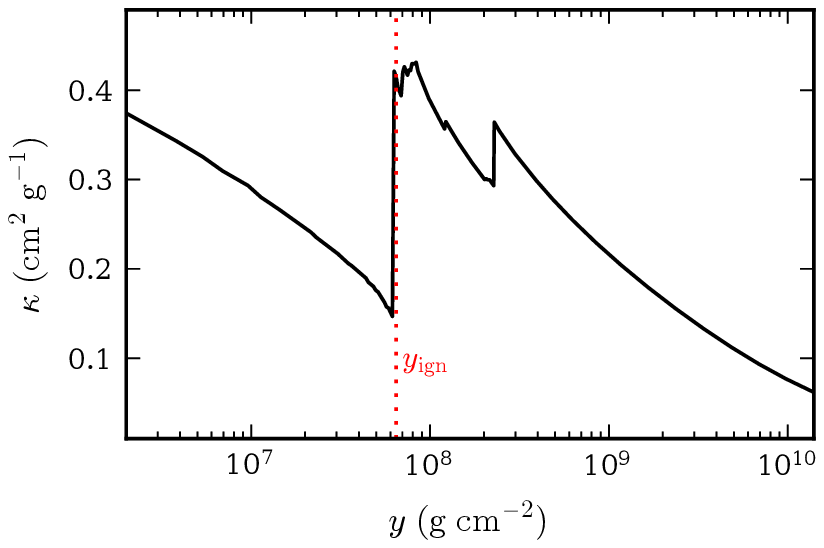}

\caption{\label{fig:kappa_y}Opacity, $\kappa$, as a function of column depth,
$y$, for the model at $t=5\,\mathrm{minutes}$ (Figure~\ref{fig:flows_xh15}).
The ignition depth, $y_{\mathrm{ign}}$, is indicated by the dotted
line. Close to this depth, there is a sharp jump in $\kappa$, from
the H-rich region at smaller depth to the H-depleted ashes at larger
depth. $\kappa$ has a maximum around $y_{\mathrm{ign}}$, and here
convection sets in. Towards larger depths, $\kappa$ exhibits small
steps at the boundaries of the different ashes layers, and the largest
step is at the iron substrate (see also Figure~\ref{fig:Composition-of-the}).}
\end{figure}
 In fact, $\kappa$ peaks in the ashes layer around $y_{\mathrm{ign}}$
(Figure~\ref{fig:kappa_y}), and there is a steep drop in $\kappa$
between the ashes layer and the hydrogen-rich layer. An increase in
$\kappa$ makes radiative cooling less efficient, and convection sets
in when $\kappa\gtrsim0.4\,\mathrm{cm^{2}\,g^{-1}}$.

As $\kappa$ peaks in the ashes layer right below the hydrogen-rich
layer, convective mixing takes place close to the left-over fuel.
Turbulent mixing can ``overshoot'' into neighboring zones, and this
brings hydrogen into the ashes layer. Some of the protons capture
onto the metals in the ashes. This produces a small peak in $\epsilon_{\mathrm{nuc}}$
and an equivalent increase in $T$, which reduces $\kappa$ (Figure~\ref{fig:flows_xh15}).
As convection is switched off, this marks the end of one mixing event.
After radiative cooling decreases $T$ and increases $\kappa$ once
again, a next convective mixing event initiates. In this way, each
burst is followed by a series of brief mixing events. Each event brings
hydrogen to larger $y$, but only when $y_{\mathrm{ign}}$ is reached,
does a full burst ignite.

After the first burst, only two mixing events were required to ignite
the second burst, but ten occurred before the third burst. The last
burst was followed by an even larger number of mixing events, but
no fourth burst appeared. One reason for this is that each subsequent
burst depletes hydrogen further from $y_{\mathrm{ign}}$ (Figure~\ref{fig:Composition-of-the},
\ref{fig:h1_mixing}), thus requiring mixing to bridge an increasing
$\Delta y$. Another aspect is the steepness of the hydrogen profile,
which is reduced by each burst (Figure~\ref{fig:Composition-of-the}).
After later bursts, the hydrogen content of the mixed-in material
is lower, and each mixing event transports hydrogen over a smaller
$\Delta y$. The mixing events after the first burst provide each
$\Delta y\simeq2.9\times10^{6}\,\mathrm{g\,cm^{-2}}$, whereas after
the second burst $\Delta y\simeq1.1\times10^{6}\,\mathrm{g\,cm^{-2}}$,
and after the third burst $\Delta y\simeq0.8\times10^{6}\,\mathrm{g\,cm^{-2}}$
(Figure~\ref{fig:h1_mixing}). The later mixing events bring in a
smaller amount of hydrogen, powering weaker burning events, which
provide smaller temperature increases, halting convection for a shorter
time. The time between subsequent mixing events is $2.9$~minutes
after burst one, $\sim0.8$~minutes after two, and $\sim0.4$~minutes
after three, respectively. 

As mentioned, during the convective episodes, a small fraction of
the mixed-in hydrogen burns by proton capture on available seed nuclei
in the ashes of the previous burst. When $y_{\mathrm{ign}}$ is reached,
however, there is a notable difference in the nuclear burning processes.
The nuclear flow through the $\beta$CNO cycle breakout reactions
$^{15}\mathrm{O}\left(\alpha,\gamma\right)\mathrm{^{19}Ne}$ and $^{18}\mathrm{Ne}\left(\alpha,p\right)\mathrm{^{21}Na}$
exceeds the $3\alpha$ flow (Figure~\ref{fig:flows_xh15}). CNO is
removed faster than it is created, while producing new seed nuclei
for the $\alpha$\textsl{p}- and \textsl{rp}-processes. This leads
to runaway burning that spreads to neighboring zones, resulting in
an SWT burst.

\subsection{Alternative Simulations: Enhanced Convection}

We perform similar simulations at $\dot{M}=0.05\,\dot{M}_{\mathrm{Edd}}$,
$\dot{M}=0.2\,\dot{M}_{\mathrm{Edd}}$, and $\dot{M}=0.3\,\dot{M}_{\mathrm{Edd}}$,
but do not find SWT bursts. This suggests that SWT bursts are confined
to a narrow range of $\dot{M}$ around $\dot{M}=0.1\,\dot{M}_{\mathrm{Edd}}$. 

\begin{figure}
\includegraphics{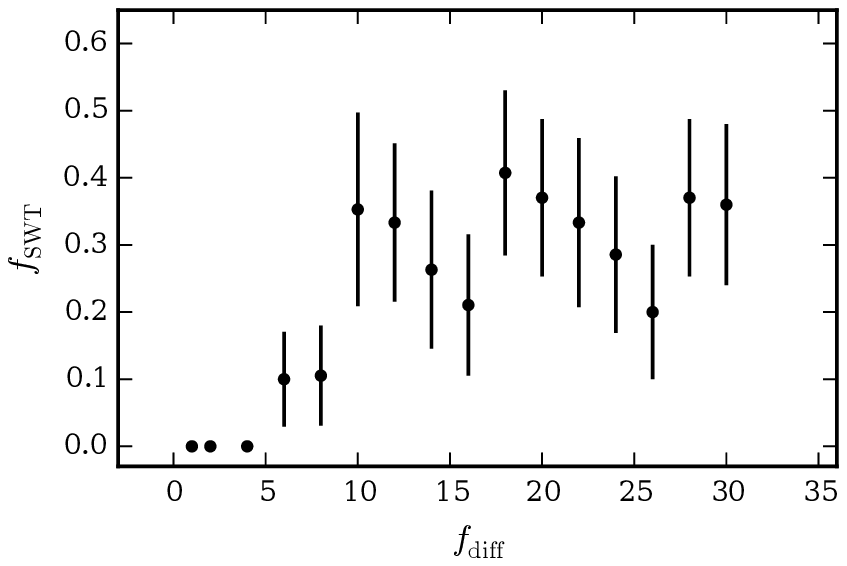}

\caption{\label{fig:grid_mdot_conv-01a}SWT fractions of models as a function
of diffusivity scale factor $f_{\mathrm{diff}}$ for $\dot{M}=0.05\dot{M}_{\mathrm{Edd}}$
and $Q_{\mathrm{b}}=0.1\,\mathrm{MeV\,u^{-1}}$.}
\end{figure}
 As an alternative to the strong base heating, we investigate the
effect of enhancing the strength of convective mixing. We assume $Q_{\mathrm{b}}=0.1\,\mathrm{MeV\,u^{-1}}$
and scale the mixing length diffusivity by a factor $f_{\mathrm{diff}}$.
This is equivalent to increasing the length scale over which convective
mixing takes place. For $\dot{M}=0.1\,\dot{M}_{\mathrm{Edd}}$ no
SWT bursts are found for $1\le f_{\mathrm{diff}}\le100$ at accretion
rates of $\dot{M}=0.1\,\dot{M}_{\mathrm{Edd}}$, $\dot{M}=0.2\,\dot{M}_{\mathrm{Edd}}$,
and $\dot{M}=0.3\,\dot{M}_{\mathrm{Edd}}$. At $\dot{M}=0.05\,\dot{M}_{\mathrm{Edd}}$,
we find SWT bursts for $f_{\mathrm{diff}}>5$, with SWT fractions
ranging from $0.1$ to $0.4$ (Figure~\ref{fig:grid_mdot_conv-01a}).
Even without strong heating, also for $Q_{\mathrm{b}}=0.1\,\mathrm{MeV\,u^{-1}}$
a substantial amount of fuel is left unburned after an LWT burst (Figure~\ref{fig:qb_overshoot}).
The distance between the left-over fuel and the ignition location
is larger for smaller $Q_{\mathrm{b}}$, but in this case the stronger
convection has a sufficiently long reach.

\section{Discussion}

\label{sec:Discussion}

SWT bursts have first been observed shortly after the discovery of
Type I bursts \citep{lewin76mnras}. Based on the relatively small
sample of bursts observed with \emph{EXOSAT}, \citet{1636:fujimoto87apj}
thought of SWT and LWT bursts as part of a continuous distribution
of recurrence times, and sought to understand them using a single
ignition mechanism. A much larger burst sample observed with the \emph{Rossi
X-ray Timing Explorer} \citep{Galloway2008catalog} and the \emph{BeppoSAX}
Wide Field Cameras \citep{Cornelisse2003} was compiled in the Multi-Instrument
Burst Archive (MINBAR), and it revealed a distinctly bimodal distribution
of $t_{\mathrm{recur}}$ \citep{Keek2010}. Furthermore, day-long
observations with \emph{XMM-Newton} and \emph{Chandra} \citep{Boirin2007,Keek2010}
showed that the bimodality is intrinsic to the source, rather than
an artifact from observing in low-Earth orbit, where Earth occultations
typically allow for at most $\sim60$~minutes of uninterrupted observation
of a source. The bimodality indicates that LWT and SWT bursts reach
ignition in a different way, which is confirmed by our simulations.
In this section we discuss the issues of left-over fuel and its reignition,
comparing previous suggested scenarios to our fully self-consistent
numerical models. 

\subsection{Left-Over Fuel}

\citet{1636:fujimoto87apj} considered the fast transport of freshly
accreted fuel down to $y_{\mathrm{ign}}$, but realized that for SWT
with $t_{\mathrm{recur}}<10$~minutes there is insufficient time
to accrete sufficient material. Unburned fuel must remain after a
burst to power an SWT burst. 

\citet{Boirin2007} noted the difficulty in explaining the recurrence
time of the LWT bursts from EXO~0748-676, as it appeared short compared
to the expected accretion time scale. They considered both hydrogen-ignited
and helium-ignited burst scenarios, and found that both explanations
required the source distance to be at either extreme of the observationally
allowed range. Assuming a larger distance results in a higher accretion
luminosity and a shorter accretion time scale. This was, however,
based on ignition models that assume the whole ignition column needs
to be replaced before the next burst. If only a fraction $f_{\mathrm{burn}}$
is consumed, the expected recurrence time is reduced by a factor $f_{\mathrm{burn}}$
as well. The fact that some fuel is left-over, therefore, also has
an observable effect on the LWT bursts.

A third observational indication for the burning of left-over fuel
is the low $\alpha$-parameter measured for SWT bursts. This parameter
is the ratio of the persistent fluence between subsequent bursts to
the burst fluence, and it provides a measure of how much is accreted
versus burned. The SWT bursts have anomalously low values of $\alpha\lesssim20$
\citep{Boirin2007}. For example, assuming the accretion of pure hydrogen
which all burns to iron in a burst, the minimum expected value is
$\alpha\simeq20$. This again indicates that the burst fuel was not
accreted within the short recurrence time. The $\alpha$ values for
the LWT bursts have typical values \citep{Boirin2007}, which may
be understood if both the persistent and burst fluence are reduced
by the same factor $f_{\mathrm{burn}}$. In our simulations we find
$\alpha$ values as low as $13$ for the SWT bursts, whereas $\alpha\ge47$
for the LWT bursts (Section~\ref{subsec:Comparison-of-LWT}).

In our series of simulations, roughly the same column of fuel remains
after each LWT burst (Figure~\ref{fig:qb_overshoot}). As these are
relatively weak bursts (the luminosity does not reach the Eddington
limit), the convection zone at the burst onset is short-lived, and
the extent of the burning region is largely set by the local conditions
at each depth \citep[see also][]{Woosley2004}. Therefore, a similar
absolute amount of fuel is left over in all models. The more important
distinction is that for models with stronger base heating (a larger
$Q_{\mathrm{b}}$), $y_{\mathrm{ign}}$ is reduced, which increases
the relative amount of unburned fuel. For the LWT bursts, $t_{\mathrm{recur}}$
is substantially lower than the expectation when all fuel burns (Figure~\ref{fig:qb_overshoot}).
Moreover, when $y_{\mathrm{ign}}$ is closer to the left-over fuel,
it is easier to transport this fuel to the location where it can ignite
to produce an SWT burst. Indeed, our simulations exhibit SWT bursts
for $Q_{\mathrm{b}}\ge2.75\,\mathrm{MeV\,u^{-1}}$. Such large values
cannot be provided by crustal heating \citep{Haensel2003,Gupta2007},
but the dissipation of rotational energy potentially is a source of
substantial heating \citep{Inogamov2010}.

\subsection{Turbulent Reignition}

\citet{Boirin2007} noted that all sources that exhibit SWT bursts
have a fast spinning neutron star, suggesting the importance of rotationally
induced processes, such as rotational mixing. Rotationally induced
mixing was also proffered by \citet{1636:fujimoto87apj} as the mechanism
to transport fuel to $y_{\mathrm{ign}}$ on a faster timescale than
the accretion timescale. The timescale for this mixing process may
indeed be similar to the SWT recurrence time \citep{Fujimoto1988A&A}.
Since this mixing process is active continuously, however, it would
produce SWT bursts all the time \citep{Piro2007}, or even stabilize
the burning \citep{Keek2009}, whereas we observe both SWT and LWT
bursts.

Our simulations do not include rotational mixing. Instead, left-over
fuel is transported to $y_{\mathrm{ign}}$ by convection. \citet{Woosley2004}
noted ``interesting episodes of convection in between the bursts.''
These convective mixing events occur in the ashes layer on a timescale
of minutes following a burst. The turbulence may extend beyond the
region where the Ledoux criterion is met, which is known as convective
overshooting. This process can grab hydrogen-rich left-over fuel and
bring it to greater depth. We find that several of such mixing events
are required to transport the fuel to $y_{\mathrm{ign}}$, where it
ignites an SWT burst (Figure~\ref{fig:h1_mixing}). The stochastic
nature of convection provides a natural explanation for the fact that
SWT bursts seemingly occur at random with a $\sim30\%$ probability.
Furthermore, the left-over fuel is mixed with the ashes, such that
the burst results from burning a diluted composition. This is consistent
with the lower hydrogen content inferred from observations \citep{Boirin2007}.
Each SWT burst depletes hydrogen to smaller depths, and a sequence
of mixing events progressively dilutes the fuel with ashes. Over time,
it is increasingly difficult to bring sufficient hydrogen down to
$y_{\mathrm{ign}}$ to ignite an SWT burst. Small mass fractions of
hydrogen burn away without triggering a runaway. This limits the number
of subsequent SWT bursts in our simulations to at most two. Afterwards,
accretion slowly replaces the burned material, and an LWT burst ignites.
The largest observed number of subsequent SWT bursts was three for
a quadruple event from 4U~1636-536 \citep{Keek2010}. The first SWT
burst was rather underluminous, which may explain why it could be
followed by two more SWT bursts.

The convective mixing events are caused by an anti-correlation of
the opacity and temperature. The free-free opacity depends on temperature
as $\kappa_{\mathrm{ff}}\propto T^{-7/2}$ \citep[e.g.,][]{Kippenhahn1994}.
The total opacity in our model, which further includes contributions
from, e.g., electron scattering, also exhibits this anti-correlation,
albeit with a slightly less steep slope (Figure~\ref{fig:kappa_t}).
During the cooling phase following a burst, the $T$ decrease leads
to an increase in $\kappa$. As radiative energy transport becomes
less efficient, convection sets in. Convection not only transports
heat outwards, but also mixes the composition inwards. The opacity
peaks in the ashes layer close to the unburned fuel. Convective overshooting,
therefore, drags hydrogen-rich material into the ashes layer. Some
of the protons are quickly captured onto seed-nuclei in the ashes.
These capture reactions increase $T$ and consequently lower $\kappa$.
Convection switches off as energy transport becomes radiative again.
Another cooling phase follows, leading up to the next convective mixing
event. The result is the periodic switching on and off of convection
on a cooling timescale of a few minutes. This behavior is reminiscent
of the $\kappa$-mechanism that produces pulsations in Cepheids, where
the anti-correlation between $T$ and $\kappa$ drives expansion and
contraction \citep[e.g.,][]{Cox1958}. The neutron star envelope acts
as a thin shell, where the strong gravitational pull of the neutron
star inhibits substantial expansion. Instead, the anti-correlation
between $T$ and $\kappa$ drives the mixing events and the burning
of hydrogen.

\subsection{Model Accuracy and Improvements}

Next, we discuss how our results depend on the implementation of our
code, in particular with respect to the opacity and convective mixing.
For the conditions relevant for the bursts in our models, the most
important contributions to the opacity are from electron scattering
and free-free transitions. They are implemented following the numerical
opacity calculations by \citet{Cox1970a,Cox1970b} (see also Section~\ref{sec:Neutron-star-envelope}).
Obtaining more accurate opacities is challenging, because of the wide
range of compositions present in the models, from solar to many different
mixtures of hydrogen, helium, and a wide range of metals. The required
mixtures are typically not covered by modern opacity tables \citep[e.g., {\scshape Opal};][]{Iglesias1996}.
\citet{Cox1965} found the differences between opacity calculations
as well as fits to calculations to be of the order $10\,\%-30\,\%$.
We, therefore, expect the uncertainties in our opacity implementation
to be at least of this size. A systematic shift in $\kappa$ influences
the conditions when convection switches on or off.

Convective mixing is implemented as a diffusive process using mixing
length theory. We find that the SWT phenomenon strongly depends on
the length scale over which mixing takes place, which is only approximated
in our model. For example, our simulations with $Q_{\mathrm{b}}=0.1\,\mathrm{MeV\,u^{-1}}$
present the largest distance between $y_{\mathrm{ign}}$ and the left-over
fuel (Figure~\ref{fig:qb_overshoot}). When the convective length
scale is increased by a factor $10$, SWT bursts appear (Figure~\ref{fig:grid_mdot_conv-01a}),
although at a lower $\dot{M}$.

Of special importance is convective overshooting. In our model, overshooting
crucially mixes hydrogen into the ashes layer. When turned off, no
SWT bursts occur. The length scale over which mixing by convective
overshooting takes place, $l_{\mathrm{mix}}$, is often assumed to
be proportional to the pressure scale height $H_{P}$, although the
precise value is highly uncertain: $l_{\mathrm{mix}}\simeq(0.1-10)H_{P}$
(e.g., \citealt{Zhang2013}). Our implementation of overshooting is,
however, crude: mixing is extended one zone beyond the boundaries
of the convective region. In tests where we reduce the zone size,
the amount of hydrogen mixed into the ashes is decreased, and the
SWT bursts disappear. Multi-dimensional hydrodynamics models are required
to improve our understanding of the length scales of convection and
convective overshooting, for example with the recently developed \textsc{Maestro}
code \citep{Malone2011,Malone2014,Zingale2015}. Methods may be developed
to use multi-dimensional simulations to inform improved implementations
of turbulent mixing in one-dimensional models \citep[e.g.,][]{Arnett2015}.
For example, one-dimensional turbulence \citep[ODT;][]{Kerstein1991}
models employ a stochastic approach.

For individual sources, SWT burst observations are restricted to a
range of mass accretion rates. This is also the case in our simulations.
This may be understood because at lower $\dot{M}$, $y_{\mathrm{ign}}$
is larger, so $f_{\mathrm{burn}}$ is higher (see Figure~\ref{fig:qb_overshoot}),
and most (all) hydrogen burns in the $\beta$CNO cycle. Therefore,
it is harder to mix hydrogen down to $y_{\mathrm{ign}}$. For the
observed sources, the upper bound coincides with a change in the state
of the accretion disk and an accompanying change in the burst regime
\citep{Galloway2008catalog,Keek2010}. Such a state transition is,
however, not part of our model. The lack of SWT bursts at $\dot{M}\ge0.2\,\dot{M}_{\mathrm{Edd}}$
needs to be investigated further, but we may find it to be due to
the approximate prescription of convective mixing and overshooting
in our model.

In summary, our models convincingly reproduce the qualitative behavior
of SWT bursts in great detail. Improvements in the implementation
of convection and convective overshooting are, however, crucial in
accurately predicting the quantitative properties of SWT bursts, including
the range of $\dot{M}$ and $Q_{\mathrm{b}}$ where they occur. 

\section{Conclusions}

\label{sec:Conclusions}

We present the first one-dimensional numerical models that include
X-ray bursts with both regular long recurrence times and recurrence
times as short as $\sim5$~minutes. The observed behavior of these
bursts is qualitatively reproduced in great detail: bursts appear
in single, double, or triple events; the SWT bursts are shorter, less
bright, and less energetic than the LWT bursts; SWT bursts occur at
$\dot{M}/\dot{M}_{\mathrm{Edd}}=0.05-0.1$; the observed fraction
of SWT bursts of $\sim30\%$ is reproduced. We find that SWT bursts
occur when after a burst some hydrogen-rich fuel is left unburned
at smaller column depth, and a series of convective mixing events
brings it down to the ignition depth, where a new burst ignites. Strong
base heating lowers the burst ignition depth, reducing its distance
from the left-over fuel, which makes it easier to produce an SWT burst.
The opacity in the ashes layer is anti-correlated with the temperature,
such that after a cooling timescale of a few minutes, the ashes become
opaque, and a convective mixing event initiates. The implementation
of convection and convective overshooting in one-dimensional codes
is rather crude. Improvements in this area are crucial to accurately
predict the mass accretion rates and amount of base heating that is
required to produce SWT bursts.

\acknowledgements{The authors are grateful to Jean in 't Zand, Tod Strohmayer, and
Simin Mahmoodifar for comments on the manuscript. L.K. is supported
by NASA under award number NNG06EO90A. The authors thank the International
Space Science Institute in Bern, Switzerland for hosting an International
Team on X-ray bursts. This work benefited from events supported by
the National Science Foundation under Grant No. PHY-1430152 (JINA
Center for the Evolution of the Elements).}

\bibliographystyle{apj}
\bibliography{short_kepler}

\end{document}